\documentclass[12pt]{article}

\begin{document}

\title{${\cal N}=1$ Supersymmetric Path-Integral Poisson-Lie Duality}

\author{ Svend E. Hjelmeland\thanks{s.e.hjelmeland@fys.uio.no}\,\,$^a$ and
 Ulf Lindstr\"om\thanks{ul@physto.se}\,\,$^{b}$
\\
\, \\
{\small{$^a$Department of Physics, University of Oslo,}}\\
{\small{P.O.Box 1048 Blindern, N-0316 Oslo, Norway}}\\
\, \\
{\small{$^{b}$ITP, University of Stockholm,}}\\
{\small{ Box 6730, S-11385 Stockholm, Sweden}}\\}

\date{\today}

\maketitle

\begin{picture}(0,0)(0,0)
\put(300,370){OSLO-TP 2-01}
\put(300,355){USITP-01-01}
\put(300,340){hep-th/0102089}
\end{picture}
\vspace{-24pt}

\begin{abstract}
We extend the path-integral formulation of Poisson-Lie duality given in
\cite{kn:TvU} to ${\cal N}=1$ supersymmetric $\sigma$-models. Using an 
explicit representation of the generators of the Drinfel'd double 
corresponding to $G\otimes U(1)^{dimG}$ we discuss an application to 
non-abelian duality. The paper also contains the relevant background and 
some comments on Poisson-Lie duality.
\end{abstract}

\newpage

\section{Introduction}
\bigskip

Poisson-Lie (PL) duality \cite{kn:KS} extends the notion of duality of 
$\sigma$-models to target spaces without isometries, replacing the isometry 
condition by the weaker PL condition. It includes ordinary abelian as
well as non-abelian duality. After its construction two obvious extensions 
seemed desirable: a quantum formulation and an application to supersymmetric 
models. 

The quantum setting has been treated in \cite{kn:AKT,kn:TvU,kn:S1}, of which 
the most relevant reference for our discussion is the path-integral treatment 
in \cite{kn:TvU}. Supersymmetric extensions have been discussed in 
\cite{kn:P,kn:S2,kn:K,kn:KP,kn:AL}.

In the present article we give an extension of the path-integral treatment 
of PL duality to ${\cal N}=1$ supersymmetric non-linear $\sigma$-models. We 
further comment on some questions of general relevance to PL duality, namely 
we give an explicit realization of the generators of the Drinfel'd double for 
non-abelian duality and we look for possible roles in PL duality for the 
extended currents that result when the Lie derivative of the background 
antisymmetric tensor field $b$ satisfies ${\cal L}b=d\omega$ \cite{kn:RV}. We 
also discuss the relation between ${\cal N}=1$ supersymmetric PL duality and 
non-abelian duality at the level of actions. To make the presentation more 
readable we have included background material on isometries, actions on group 
manifolds, non-abelian duality and PL duality. An introduction on duality 
which may serve as further background is found in \cite{kn:GR1,kn:GR2}.

The outline of the paper is the following. We start by giving some background 
material on isometries and actions on group manifolds. Then in 
sect.\ref{sectNAD} we give the basics of non-abelian duality. 
In sect.\ref{sectPLD} we give a short summary of PL duality including 
the Drinfel'd double and the path-integral derivation of Tyurin and von Unge 
\cite{kn:TvU}. In sect.\ref{sectSPLD} we generalize this path-integral 
to ${\cal N}=1$ supersymmetry. The non-abelian limit of the
${\cal N}=1$ is presented in sect.\ref{sectSNAD} in where we explicitly 
construct a set of generators spanning the Drinfel'd double 
$G\otimes U(1)^{dimG}$, where $G$ must be a compact group. We also find the 
gauge fixed action in the non-abelian limit. In two appendices we discuss
a modified current as well as $WZW$ models in the PL setting.

\section{Actions and isometries}
\label{sect2}

A string propagating in a curved background with metric $g_{ij}$, $NS$ 
antisymmetric tensor field $b_{ij}$ and dilaton $\phi$, has a bosonic part 
given by the $\sigma$-model action
\begin{eqnarray}
\label{stringaction}
S&=&-\frac{1}{4\pi\alpha'}\int_{W} d^2\sigma\left(\sqrt{-h}h^{mn}
\partial_{m}x^{i}\partial_{n}x^{j}g_{ij}(x^{k})+
\epsilon^{mn}\partial_{m}x^{i}\partial_{n}x^{j}b_{ij}(x^{k})
\right. \nonumber \\ 
&&\left.
\qquad\qquad\qquad\quad -\alpha'\sqrt{-h}R^{(2)}\phi(x^{k})\right),
\end{eqnarray}
where $x^{i}$ ($i=1,\ldots,\mbox{dim}T$) are the coordinates on the target 
space $T$ (spacetime), $h_{mn}$ is the auxiliary world-sheet metric, and 
$R^{(2)}$ the corresponding curvature.
With application to strings in mind we thus study the $\sigma$-model
(in units where $\alpha'=1/2\pi$)
\begin{equation}
\label{boseaction1}
S[x]=\int_{W}d^{2}\xi\partial x^{i}(g_{ij}+b_{ij})(x)\bar{\partial}x^{j}=
\int_{W}d^{2}\xi\partial x^{i}f_{ij}(x)\bar{\partial}x^{j},
\end{equation}
which is (\ref{stringaction}) in conformal gauge, light-cone coordinates 
$\xi=\frac{1}{2}(\tau+\sigma)$ and $\bar{\xi}=\frac{1}{2}(\tau-\sigma)$ and 
with the dilaton field $\phi$ set to zero (later we comment briefly on 
non-zero $\phi$).

We shall be interested in the case when (some of) the $x^{i}$'s are acted on
by a Lie-group $G$ with Lie-algbra ${\cal G}$ generated by $\{t_{a}\}$
\begin{equation}
[t_{a},t_{b}]=f_{ab}^{~~c}t_{c}.
\end{equation}
The relevant part of (\ref{boseaction1}) may then be written as
\begin{equation}
\label{boseaction2}
S[g]=\int_{W}d^{2}\xi(g^{-1}\partial g)^{a}E_{ab}(g)
(g^{-1}\bar{\partial}g)^{b},
\end{equation}
where the correspondence is via the left and right invariant forms on 
${\cal G}$ 
\begin{equation}
\label{LR-former}
L=g^{-1}dg=t_{a}L^{a}_{~i}dx^{i};\ \ \ \ \ 
R=dgg^{-1}=t_{a}R^{a}_{~i}dx^{i}.
\end{equation}
The coordinates $x^{i}$ parametrize the group manifold ${\cal M}$, the group
action is
\begin{equation}
\label{graction}
\delta x^{i}=\epsilon^{a}R^{i}_{a},
\end{equation}
and
\begin{equation}
\label{backgr}
E_{ab}=L^{i}_{a}f_{ij}L^{j}_{b}.
\end{equation}
For the special case of interest for strings when (\ref{stringaction}) 
corresponds to a $WZW$-model, the correspondence is
\begin{equation}
\label{wzw}
S[g]=\int_{\partial Y} d^2\xi Tr(g^{-1}\partial g)(g^{-1}\bar{\partial}g)+
\frac{1}{3}\int_Y Tr(g^{-1} dg\wedge g^{-1} dg\wedge g^{-1} dg),
\end{equation}
with
\begin{equation}
\label{metrikk}
g_{ij}=L^{a}_{i}\eta_{ab}L^{b}_{j}=R^{a}_{i}\eta_{ab}R^{b}_{j},
\end{equation}
and $\eta_{ab}=Tr(t_{a}t_{b})$ the Cartan-Killing metric. Further, the 
torsion is in this case given by the left or right invariant forms as
$H_{ijk}=\frac{1}{2}L_{i}^{a}L_{j}^{b}L_{k}^{c}f_{abc}=
\frac{1}{2}R_{i}^{a}R_{j}^{b}R_{k}^{c}f_{abc}$
which gives the $"x"$-version of the last term in (\ref{wzw}) as
\begin{equation}
\label{wzw-term}
S_{WZW}[x]=\frac{1}{3!}\int_{Y}d^{3}y
\epsilon^{\mu\nu\lambda}H_{ijk}\partial_{\mu}x^{i}\partial_{\nu}x^{j}
\partial_{\lambda}x^{k}=
\int_{W=\partial Y}\!\!\!\!\!\!\!\!\!d^{2}\xi\partial x^{i}b_{ij}
\bar{\partial}x^{j},
\end{equation}
where we assume conditions such that the last integral is well-defined 
\cite{kn:HS}.

As a first example of a group $G$ we consider (generalized) isometries of
the $\sigma$-model (\ref{boseaction1}). We thus take the right-invariant
forms (say) to be Killing vectors $R_{a}^{i}=k^{i}_{a}$ and the algebra is
$[k_{a},k_{b}]=f_{ab}^{~~c}k_{c}$. The action (\ref{boseaction1}) or
(\ref{wzw}) is now invariant under the group action (\ref{graction}), 
$\delta x^{i}=\epsilon^{a}k^{i}_{a}$, due to the generalized isometry 
conditions
\begin{equation}
\label{iso-cond1}
{\cal L}_{k_{a}}g_{ij}=0,\ \ \ \ \
{\cal L}_{k_{a}}H_{ijk}=0.
\end{equation}
The latter of these is equivalent to \cite{kn:RV}
\begin{equation}
\label{iso-cond2}
{\cal L}_{k_{a}}b_{ij}=\partial_{[i}\omega_{j]a},
\end{equation}
for some one-forms $\omega_{ia}$. Note that $b_{ij}$ is defined only up to 
(spacetime) gauge transformations 
$b_{ij}\rightarrow b_{ij}+\partial_{[i}\lambda_{j]}$, where $\lambda_{j}$ are
the components of some one-form $\lambda$. Under this gauge transformation the
$\omega_{ai}$ transform as 
$\omega_{ai}\rightarrow\omega_{ai}+{\cal L}_{k_{a}}\lambda_{i}$.

For the $WZW$-model the $WZW$-term (\ref{wzw-term}) varies into
\begin{equation}
\label{varyWZW}
\delta S_{WZW}=\frac{1}{2!}\int_{W}d^{2}\sigma
\epsilon^{a}k^{i}_{a}H_{ijk}\partial_{\mu}x^{j}
\partial_{\nu}x^{k}\epsilon^{\mu\nu}.
\end{equation}
This integral vanishes (up to a surface term) if $k^{i}_{a}H_{ijk}=
\partial_{[j}v_{k]a}$ \cite{kn:HS}, where $v_{ia}$ is a component of a 1-form.
This is again satisfied under (\ref{iso-cond1}) and (\ref{iso-cond2}).

In general the Noether currents corresponding to the symmetries generated by 
$k_{a}^{i}$
are \cite{kn:RV}
\begin{equation}
\label{mod-curr}
\hat{J}_{a}=\partial x^{i}f_{ij}k^{j}_{a}+\omega_{ai}\partial x^{i};\ \ \ \ \
\hat{\bar{J}}_{a}=k^{i}_{a}f_{ij}\bar{\partial}x^{j}-
\omega_{ai}\bar{\partial}x^{i}.
\end{equation}
The equation of motion state that this current is 
conserved
\begin{equation}
\partial\hat{\bar{J}}_{a}+\bar{\partial}\hat{J}_{a}=0.
\end{equation}

\section{Non-abelian dualization}
\label{sectNAD}

When a $\sigma$-model has generalized isometries as described above it may be 
dualized \cite{kn:GR2,kn:FQ}. In this section we give a brief description of 
this non-abelian dualization. First the fields $x^{i}$ in (\ref{boseaction1}) 
are separated into those acted on by the isometry $x^{\hat{i}}$ and those not 
acted on $x^{\alpha}$ (spectators)
\begin{equation}
\label{bosespectators}
S[x^{\hat{i}},x^{\alpha}]=\int d^{2}\xi
\left[\partial x^{\hat{i}}f_{\hat{i}\hat{j}}\bar{\partial}x^{\hat{j}}
+\partial x^{\hat{i}}f_{\hat{i}\beta}\bar{\partial}x^{\beta}
+\partial x^{\alpha}f_{\alpha\hat{j}}\bar{\partial}x^{\hat{j}}
+\partial x^{\alpha}f_{\alpha\beta}\bar{\partial}x^{\beta}\right].
\end{equation}
Secondly, the isometry
\begin{equation}
\label{localkilling}
\delta x^{\hat{i}}=
\epsilon^{a}(\xi,\bar{\xi})k^{\hat{i}}_{a}(x^{\hat{j}},x^{\alpha})
\end{equation}
is gauged. The gauging of the symmetric part of the action 
(\ref{bosespectators}) by 
minimal coupling is straight forward, but to gauge the antisymmetric part by 
minimal coupling the torsion potential must be well-defined and 
Lie-invariant (i.e. ${\cal L}_{k_{a}}b_{ij}=0$) \cite{kn:HS}. Letting 
$\partial x^{\hat{i}}\rightarrow \partial x^{\hat{i}}+A^{a}L^{\hat{i}}_{a}$ and
$\bar{\partial}x^{\hat{i}}\rightarrow 
\bar{\partial}x^{\hat{i}}+\bar{A}^{a}L^{\hat{i}}_{a}$ gives the gauged 
version of (\ref{bosespectators}). 
A first order action is found by fixing a gauge 
$(\partial x^{\hat{i}}=\bar{\partial}x^{\hat{i}}=0)$, and adding a term 
including Lagrange multipliers 
\begin{eqnarray}
\label{firstorder}
S^{(1)}[x,A,\lambda]&=&\int d^{2}\xi
\left[A^{a}E_{ab}\bar{A}^{b}+A^{a}F^{R}_{a\beta}\bar{\partial}x^{\beta}
+\partial x^{\alpha}F^{L}_{\alpha b}\bar{A}^{b}
+\partial x^{\alpha}f_{\alpha\beta}\bar{\partial}x^{\beta}\right] \nonumber \\
&&\qquad\quad+\int d^{2}\xi\lambda_{a}f^{a},
\end{eqnarray}
where we have defined 
$E_{ab}=L^{\hat{i}}_{a}f_{\hat{i}\hat{j}}L^{\hat{j}}_{b}$, 
$F^{R}_{a\beta}=L^{\hat{i}}_{a}f_{\hat{i}\beta}$ and 
$F^{L}_{\alpha b}=f_{\alpha\hat{j}}L^{\hat{j}}_{b}$.
Here $\lambda_{a}$ are Lagrange multipliers that take values in 
the Lie-algebra ${\cal G}$ and transforms in the adjoint representation. The 
field strength is 
$f^{a}=\partial\bar{A}^{a}-\bar{\partial}A^{a}+f^{a}_{bc}A^{b}\bar{A}^{c}$
where the gauge-fields $A^{a}$, $\bar{A}^{a}$ take values in the Lie algebra 
of the isometry group.

Varying the first-order action w.r.t. $\lambda_{a}$ one obtains, at least
locally, $A^{a}=(g^{-1}\partial g)^{a}$, 
$\bar{A}^{a}=(g^{-1}\bar{\partial} g)^{a}$. 
So modulo global issues \cite{kn:GR2} the 
first order action goes into the original one when substituting this solution.
If one instead integrates out
the gauge fields $A^{a}$, $\bar{A}^{a}$ one should get the ``dual'' action. 
However, the symmetry of the background that is used in the abelian case
to dualize back to the original action is non-local
in the non-abelian case. Dualization of this non-local symmetry does 
not recover the original action \cite{kn:GR2}.

In the abelian case where $k^{i}_{a}$ commute the original $f_{ij}$ and dual 
$\tilde{f}_{ij}$ backgrounds are related 
via the Buscher transformation \cite{kn:BT}
\begin{eqnarray}
\label{abelianBT}
\tilde{f}_{\hat{i}\hat{j}}&=&(f^{-1})_{\hat{i}\hat{j}};\ \ \ \ \
\tilde{f}_{\hat{i}\beta}=(f^{-1})_{\hat{i}}^{~\hat{j}}f_{\hat{j}\beta}; 
\nonumber \\
\tilde{f}_{\alpha\hat{j}}&=&-f_{\alpha\hat{i}}(f^{-1})^{\hat{i}}_{~\hat{j}};
\ \ \ \ \
\tilde{f}_{\alpha\beta}=f_{\alpha\beta}-
f_{\alpha\hat{i}}(f^{-1})^{\hat{i}\hat{j}}f_{\hat{j}\beta}.
\end{eqnarray} 
The transformation maps manifolds without torsion 
($f_{\hat{i}\beta}=f_{\beta\hat{i}}$) on to manifolds with torsion 
($\tilde{f}_{\hat{i}\beta}=-\tilde{f}_{\beta\hat{i}}$).

\section{Poisson-Lie duality}
\label{sectPLD}

A more general and useful scheme for finding dual actions,
not based on the existence of generalized isometries, is the 
PL duality \cite{kn:KS} where the 
isometry is replaced with a weaker condition. This duality is most easily 
discussed in terms of the would-be Noether currents of the transformations 
$\delta x^{i}=\epsilon^{a}R^{i}_{a}$;
\begin{equation}
\label{curr}
J_{a}=\partial x^{i}f_{ij}R^{j}_{a};\ \ \ \ \ 
\bar{J}_{a}=R^{i}_{a}f_{ij}\bar{\partial}x^{j}.
\end{equation}
Since ${\cal L}_{R_{a}}f_{ij}\neq 0$ these transformations are no-longer a 
symmetry of the action (for constant $\epsilon^{a}$), in fact
\begin{equation}
\label{varS}
\delta S=\int d^{2}\xi\partial x^{i}\bar{\partial}x^{j}
\epsilon^{a}({\cal L}_{R_{a}}f_{ij}),
\end{equation}
and the field equations
\begin{equation}
\label{condition}
\partial\bar{J}_{a}+\bar{\partial}J_{a}-
{\cal L}_{R_{a}}f_{ij}\partial x^{i}\bar{\partial}x^{j}=0,
\end{equation}
no-longer look like Bianchi identities. However, Klim\v{c}ik and 
\v{S}evera \cite{kn:KS} introduced the following generalization of the isometry
condition for the background
\begin{equation}
\label{condition1}
{\cal L}_{R_{a}}f_{\hat{i}\hat{j}}=
-f_{\hat{i}\hat{k}}R^{\hat{k}}_{b}\tilde{f}^{bc}_{a}
R^{\hat{l}}_{c}f_{\hat{l}\hat{j}},
\end{equation}
which turns (\ref{condition}) into
\begin{equation}
\label{Maurer1}
\partial\bar{J}_{a}+\bar{\partial}J_{a}+J_{b}\tilde{f}^{bc}_{a}\bar{J}_{c}=0.
\end{equation}
Here $\tilde{f}^{ab}_{c}$ are structure constants in a dual Lie algebra, and
Klim\v{c}ik and \v{S}evera went on to show that the condition 
(\ref{condition}) can be solved
and the dual model found provided that the Lie algebra ${\cal G}$ and its dual
$\tilde{\cal G}$ form what is called a Drinfel'd double 
\cite{kn:D,kn:FG,kn:AM} which we now briefly decribe.

Let $G$ and $\tilde{G}$ be symmetry groups of the original $\sigma$-model
and the dual one, respectively, with $\mbox{dim}G=\mbox{dim}\tilde{G}$. 
The corresponding Lie algebras are ${\cal G}$ and $\tilde{\cal G}$.
Then the Drinfel'd double $D\equiv G\otimes\tilde{G}$ and
comes equipped with an invariant inner product $\langle~,~\rangle$ 
and the corresponding algebra ${\cal D}$ 
consists of the two subalgebras 
${\cal G}$ and $\tilde{\cal G}$ that are null-spaces w.r.t. this product. 
We choose two sets of generators $\{T_{a}\}$ and $\{T^{a}\}$ so that
$\{T_{a}\}$ span ${\cal G}$ and $\{T^{a}\}$ span $\tilde{\cal G}$.
The set $T_{A}\in\{T_{a},T^{b}\}$ then span ${\cal D}$. The Lie algebra of 
the Drinfel'd double generated by $T_{a}$ and $T^{a}$ 
$(a=1,\ldots,\mbox{dimG})$, is 
\begin{eqnarray}
\label{algebra}
[T_{a},T_{b}]&=&f_{ab}^{c}T_{c}, \nonumber \\
~[T^{a},T^{b}]&=&\tilde{f}^{ab}_{c}T^{c}, \nonumber \\
~[T_{a},T^{b}]&=&\tilde{f}^{bc}_{a}T_{c}-f_{ac}^{b}T^{c},
\end{eqnarray}
where $f_{ab}^{c}$ and $\tilde{f}^{ab}_{c}$ are the structure constants of 
${\cal G}$ and $\tilde{\cal G}$, respectively, and satisfy
the bi-Lie algebra $({\cal G},\tilde{\cal G})$ consistency condition 
\begin{equation}
\label{consistency}
f^{a}_{dc}\tilde{f}^{rs}_{a}=
\tilde{f}^{as}_{c}f^{r}_{da}+\tilde{f}^{ra}_{c}f^{s}_{da}
-\tilde{f}^{as}_{d}f^{r}_{ca}-\tilde{f}^{ra}_{d}f^{s}_{ca}.
\end{equation}
This condition arises in the PL duality context as the condition
$[{\cal L}_{k_{a}},{\cal L}_{k_{b}}]=f_{ab}^{~~c}{\cal L}_{k_{c}}$ applied
to (\ref{condition1}).
The invariant inner product between the generators has the following properties
\begin{equation}
\label{innerprod}
\langle T_{a},T_{b}\rangle =\langle T^{a},T^{b}\rangle =0,
~~~~~~\langle T_{a},T^{b}\rangle =\delta_{a}^{~b}
\end{equation}
and obeys the invariance condition
\begin{equation}
\langle {\cal X}T^{A}{\cal X}^{-1}, T^{B}\rangle =
\langle T^{A},{\cal X}^{-1}T^{B}{\cal X}\rangle,
\end{equation} 
where ${\cal X}$ is any element of the Drinfel'd double or one of its 
subgroups.

We define,
\begin{eqnarray}
\label{def}
\mu^{ab}(g)&=&\langle gT^{a}g^{-1},T^{b}\rangle ;\ \ \ \ \
\nu^{a}_{~b}(g)=\langle gT^{a}g^{-1},T_{b}\rangle; \nonumber \\
\alpha_{b}^{~a}(\tilde{g})
&=&\langle \tilde{g}T_{b}\tilde{g}^{-1},T^{a}\rangle ;\ \ \ \ \
\beta_{ab}(\tilde{g})=\langle \tilde{g}T_{a}\tilde{g}^{-1},T_{b}\rangle
\end{eqnarray}
which obey $\mu(g^{-1})=\mu^{t}(g)$, $\nu(g^{-1})=\nu^{-1}(g)$, 
$\alpha(\tilde{g}^{-1})=\alpha^{-1}(\tilde{g})$ 
and $\beta(\tilde{g}^{-1})=\beta^{t}(\tilde{g})$ where $t$ stands for 
transpose.

We now return to the solution of (\ref{condition1}) given by Klim\v{c}ik and 
\v{S}evera. With $f_{\hat{i}\hat{j}}=L^{a}_{\hat{i}}E_{ab}L^{b}_{\hat{j}}$ as 
in (\ref{backgr}) the solution is
\begin{equation}
E_{ab}=((E^{0})^{-1}+\Pi)^{-1}_{ab};\ \ \ \ \ 
\Pi^{ab}=\mu^{ac}\nu_{c}^{~b}.
\end{equation}
Similarly, in the dual theory one has relations corresponding to 
(\ref{condition1}) and (\ref{Maurer1}) and
\begin{equation}
\tilde{E}^{ab}=[(E^{0}+\tilde{\Pi})^{-1}]^{ab};
\ \ \ \ \ \ \tilde{\Pi}_{ab}=\beta_{ac}\alpha^{c}_{~b}.
\end{equation}
The abelian and non-abelian dualities described previously are
special cases of the more general PL duality. In the non-abelian
case we have $\mu^{ab}=0$, $\alpha^{a}_{~b}=\delta^{a}_{~b}$ and
$\beta_{ab}=f^{c}_{ab}\tilde{x}_{c}$, where $\tilde{x}_{c}$ is the dual 
non-inert coordinates, so that $E_{ab}=E^{0}_{ab}$ and 
$\tilde{E}^{ab}=[(E^{0}+f^{c}\tilde{x}_{c})^{-1}]^{ab}$.

As described above PL duality acts at the classical
level. An important step towards the quantum implementation was taken
in \cite{kn:TvU}, where it was shown how PL duality can be 
derived 
from a constrained $WZW$-model defined on the Drinfel'd double $D$,
starting from the path-integral
\begin{equation}
\label{PIbose}
{\cal Z}=\int{\cal D}l{\cal D}x
\delta[\langle l^{-1}\partial l,T^{a}\rangle E^{0}_{ab}+
\partial x^{\alpha}F^{L}_{\alpha b}-\langle l^{-1}\partial l,T_{b}\rangle]
e^{-I[l,x]},
\end{equation}
where
\begin{equation}
I[l,x]=I[l] +
\int d^{2}\xi[\langle l^{-1}\partial l,T^{a}\rangle F^{R}_{a\alpha}
\bar{\partial}x^{\alpha}+
\partial x^{\alpha}\hat{f}_{\alpha\beta}\bar{\partial}x^{\beta}],
\end{equation}
and $l\in D$.
It should be stressed that $\hat{f}_{\alpha\beta}$ is generally not equal to 
$f_{\alpha\beta}$ that appears in (\ref{firstorder})\footnote{In \cite{kn:TvU}
this is not clear.}. The WZW-model $I[l]$ on the Drinfel'd double can be 
written
\begin{equation}
I[l]=\int d^{2}\xi\langle l^{-1}\partial l,l^{-1}\bar{\partial}l\rangle
+\int d^{3}y\langle l^{-1}\partial_{t} l,
[l^{-1}\partial l,l^{-1}\bar{\partial}l]\rangle.
\end{equation}
We also note that the path-integral (\ref{PIbose}) can
be obtained from the path-integral
\begin{equation}
{\cal Z}=\int{\cal D}l{\cal D}x{\cal D}\bar{c}e^{-I[l,x,\bar{c}]},
\end{equation}
where
\begin{equation}
\label{I+Lagrmult}
I[l,x,\bar{c}]=I[l,x]+\int d^{2}\xi[\langle l^{-1}\partial l,T^{a}
\rangle E^{0}_{ab}+
\partial x^{\alpha}F^{L}_{\alpha b}
-\langle l^{-1}\partial l,T_{b}\rangle]\bar{c}^{b}
\end{equation}
by integrating out the Lagrange multiplier $\bar{c}^{b}$ (We use 
(\ref{I+Lagrmult}) in section \ref{sectSNAD} where we study the 
${\cal N}=1$ supersymmetric generalization of non-abelian duality).

The dualization process goes as follows. The original action is recovered when
the group element $l$ in (\ref{PIbose}) is decomposed as $l=\tilde{h}g$. 
This makes the left-invariant Haar measure split into 
${\cal D}(\tilde{h}g)=
{\cal D}\tilde{h}{\cal D}g\det(\nu^{-1})$, where $\nu\equiv\nu^{a}_{~b}$ is
defined in eq.(\ref{def}). By letting $Tr\rightarrow\langle~,~\rangle$
in the Polyakov-Wiegmann formula \cite{kn:PW}
\begin{equation}
\label{PW}
S[g_1 g_2]=S[g_1]+S[g_2]+\int d^{2}\xi
Tr(g^{-1}_{1}\partial g_{1}\bar{\partial}g_{2}g_{2}^{-1})
\end{equation}
and using the inner product defined in (\ref{innerprod}) the two first terms
of $I[\tilde{h}g]$, namely $I[\tilde{h}]$ and $I[g]$, are zero.
Integrating out $\tilde{h}$ from the remaining part of $I[\tilde{h}g]$
gives back the original action (\ref{bosespectators}). When we decompose $l$ 
as $l=h\tilde{g}$ and integrate out $h$ we obtain the dual action. 

This procedure gives non-trivial Jacobians that must be 
regularized in the quantum theory. Using heat kernel regularization, it was 
shown in \cite{kn:TvU} that as a result the dilaton in both the original and 
the dual theory gets an extra shift. In the original theory the shift is
\begin{equation}
\label{dilatontransf1}
\phi=\phi^{0}+\ln\det E(g,x^{\alpha})-\ln\det E^{0}(x^{\alpha})
\end{equation}
while in the dual theory it transforms as
\begin{equation}
\label{dilatontransf2}
\tilde\phi=\phi^{0}+\ln\det\tilde{E}(\tilde{g},x^{\alpha}).
\end{equation}
In the non-abelian limit when $E(g,x^{\alpha})=E^{0}(x^{\alpha})$ we see that 
we have the usual 
transformation law between $\phi$ and $\tilde\phi$, namely
\begin{equation}
\tilde\phi=\phi+\ln\det\tilde{E}(\tilde{g},x^{\alpha}),
\end{equation}
where $\tilde{E}^{ab}=[(E^{0}+f^{c}\tilde{x}_{c})^{-1}]^{ab}$.

\section{${\cal N}=1$ Supersymmetric Poisson-Lie Duality}
\label{sectSPLD}

In this section we present PL duality in a supersymmetric setting.
For $(1,0)$ and $(1,1)$ supersymmetry this is discussed at the classical level
in \cite{kn:K}. The path-integral formulation is new.

The presence of one supersymmetry on the world-sheet introduces additional 
spinors that transform under the action of the Drinfel'd double but otherwise
changes very little. This is similar to the non-abelian duality, where the
gauging of isometries needed follow essentially the bosonic pattern for
${\cal N}=1$ \cite{kn:HKLR}.
Hence, the presentation of the basics of the Drinfel'd double given above
can be taken over to the ${\cal N}=1$ case. However the elements of the
Drinfel'd double $D$ and its subgroups $G$ and $\tilde{G}$ are now real
${\cal N}=1$ superfields.  

The ${\cal N}=1$ generalization of the bosonic action in (\ref{boseaction1}) 
is\footnote{We write the algebra as $\{D_{A},D_{B}\}=2i\partial_{AB}$; 
$\partial_{AB}=(\gamma^{m})_{AB}\partial_{m}$, where 
$(\gamma^{m})_{A}^{~~B}=(\sigma^{2},-i\sigma^{1})$. We further define
$(\gamma)_{A}^{~~B}=\sigma^{3}$ and we raise and lower indices
using $C_{AB}=-C^{AB}=\sigma^{2}$.} \cite{kn:GHR}
\begin{equation}
\label{susyaction1}
S=-\frac{1}{4}\int d^{2}\sigma d^{2}\theta
\left[G_{ij}(X)D^{A}X^{i}D_{A}X^{j}
-B_{ij}(X)D^{A}X^{i}(\gamma D)_{A}X^{j}\right],
\end{equation}
where $A,B$ are spinor indices, $G_{ij}$ and $B_{ij}$ are the metric and the 
torsion potential, respectively, and $X^{i}$ 
($i=1,\ldots ,\mbox{dim${T}$}$) are real scalar superfields.
The $\theta$-independent components of $X^{i}$ is $x^{i}$.
The $\theta$-independent components of $G_{ij}$ and $B_{ij}$ are
$g_{ij}$ and $b_{ij}$, respectively. 
For our discussion it is convenient to rewrite 
the action using the $\pm$-notation. Seperating out the spectators 
$(X^{\alpha})$
we get
\begin{eqnarray}
\label{act.w.spect}
S&=&i\int d^{2}\xi d^{2}\theta 
\left[D_{+}X^{\hat{i}}F_{\hat{i}\hat{j}}D_{-}X^{\hat{j}}+
D_{+}X^{\hat{i}}F_{\hat{i}\beta}D_{-}X^{\beta}+
D_{+}X^{\alpha}F_{\alpha\hat{j}}D_{-}X^{\hat{j}}
\right. \nonumber \\
&&\qquad\qquad\qquad + \left.
D_{+}X^{\alpha}F_{\alpha\beta}D_{-}X^{\beta}\right],
\end{eqnarray}
where $F_{ij}=G_{ij}+B_{ij}$ and $i=(\hat{i},\alpha)$. 

Again we want to study the case when the $x^{\hat{i}}$ transform under some 
group $G$, and thus study the $\sigma$-model on some group manifold ${\cal M}$.
The fields are ${\cal N}=1$ real scalar group-valued superfield, $U\in G$, 
whose $\theta$-independent part is $g$, and the spectators $X^{\alpha}$ which 
are ${\cal N}=1$ real superfields. The corresponding action on ${\cal M}$ may 
be written
\begin{eqnarray}
\label{gr.act.w.spect}
S&=&i\int d^{2}\xi d^{2}\theta
\left[(U^{-1}D_{+}U)^{a}E_{ab}(U^{-1}D_{-}U)^{b}+
(U^{-1}D_{+}U)^{a}F^{R}_{a\beta}D_{-}X^{\beta}
\right. \nonumber \\
&&\qquad\qquad\qquad + \left.
D_{+}X^{\alpha}F^{L}_{\alpha b}(U^{-1}D_{-}U)^{b}+
D_{+}X^{\alpha}F_{\alpha\beta}D_{-}X^{\beta}\right].
\end{eqnarray}
To make contact between (\ref{act.w.spect}) and (\ref{gr.act.w.spect}) we 
follow the same route as in the bosonic 
case. The superfields $X^{\hat{i}}$ in (\ref{act.w.spect})  are thus taken 
to transform under $G$ and are related to $U$ via left invariant one-form 
$U^{-1}D_{\pm}U=T_{a}L^{a}_{~\hat{i}}D_{\pm}X^{\hat{i}}$. 
Varying the action under
\begin{equation}
\label{var}
\delta X^{\hat{i}}=
\epsilon^{a}(\xi,\bar{\xi},\theta)R^{\hat{i}}_{a}(X^{\hat{j}},X^{\alpha})
\end{equation}
gives the field equations
\begin{equation}
\label{fespect}
D_{(+}J_{a-)}-\sum_{i=(\hat{i},\alpha)}
{\cal L}_{R_{a}}F_{ij}D_{+}X^{i}D_{-}X^{j}=0,
\end{equation}
where the ``currents''
\begin{equation}
J_{a+}=-(D_{+}X^{\hat{i}}F_{\hat{i}\hat{j}}
+D_{+}X^{\alpha}F_{\alpha\hat{j}})R_{a}^{\hat{j}};\ \ \ \ \
J_{a-}=R_{a}^{\hat{i}}(F_{\hat{i}\hat{j}}D_{-}X^{\hat{j}}
+F_{\hat{i}\beta}D_{-}X^{\beta}),
\end{equation}
are Noether currents when $R^{\hat{i}}_{a}$ generates an isometry.
For these currents to satisfy 
a Maurer-Cartan equation we again introduce a PL condition. 
The supersymmetric version of the Maurer-Cartan equation (\ref{Maurer1}) is
\begin{equation}
\label{susyMC}
D_{(+}J_{a-)}-J_{b+}\tilde{f}^{bc}_{a}J_{c-}=0,
\end{equation}
and the PL condition for the background that ensures that 
(\ref{fespect}) is equivalent to (\ref{susyMC}) are
\begin{eqnarray}
\label{KlimSev}
{\cal L}_{R_{a}}F_{\hat{i}\hat{j}}&=&
-F_{\hat{i}\hat{k}}R^{\hat{k}}_{b}\tilde{f}^{bc}_{a}
R^{\hat{l}}_{c}F_{\hat{l}\hat{j}}; \ \ \ \ \
{\cal L}_{R_{a}}F_{\hat{i}\beta}=
-F_{\hat{i}\hat{k}}R^{\hat{k}}_{b}\tilde{f}^{bc}_{a}
R^{\hat{l}}_{c}F_{\hat{l}\beta}; \nonumber \\
{\cal L}_{R_{a}}F_{\alpha\hat{j}}&=&
-F_{\alpha\hat{k}}R^{\hat{k}}_{b}\tilde{f}^{bc}_{a}
R^{\hat{l}}_{c}F_{\hat{l}\hat{j}}; \ \ \ \ \
{\cal L}_{R_{a}}F_{\alpha\beta}=
-F_{\alpha\hat{j}}R^{\hat{j}}_{b}\tilde{f}^{bc}_{a}
R^{\hat{l}}_{c}F_{\hat{l}\beta},
\end{eqnarray}
which is the supersymmetric version of (\ref{condition1}). 

Now we want to extend the analysis given in \cite{kn:TvU} to ${\cal N}=1$
supersymmetry. A direct generalization of (\ref{PIbose}) to ${\cal N}=1$ 
supersymmetry is the following constrained ${\cal N}=1$ functional
\begin{equation}
\label{PIsusy}
{\cal Z}=\int{\cal D}L{\cal D}X
\delta[\langle L^{-1}D_{+}L,T^{a}\rangle E^{0}_{ab}
+D_{+}X^{\alpha}F^{L}_{\alpha b}-\langle L^{-1}D_{+}L,T_{b} \rangle]
\exp(-I[L,X]),
\end{equation}
where the action
\begin{equation}
\label{susyaction2}
I[L,X]=I[L]+i\int d^{2}\xi d^{2}\theta
\left[
\langle L^{-1}D_{+}L,T^{a}\rangle F^{R}_{a\beta}D_{-}X^{\beta}
+D_{+}X^{\alpha}\hat{F}_{\alpha\beta}D_{-}X^{\beta}\right],
\end{equation}
and where $I[L]$ is an ${\cal N}=1$ WZW model on the Drinfel'd double.
Here $E^{0}_{ab}$, $F^{L}_{\alpha b}$, $F^{R}_{a\beta}$ and 
$\hat{F}_{\alpha\beta}$ depend on the spectator superfields $X^{\alpha}$.

The element $L\in D$ in (\ref{PIsusy}) can be decomposed near 
the identity in two ways;
$L=\tilde{V}U=V\tilde{U}$, where $U\in G$ and $\tilde{V}\in\tilde{G}$. To 
derive the original action we decompose
$L$ as $\tilde{V}U$. After integrating out $\tilde{V}$ via a change of
variables we can read off the metric and the torsion potential:
\begin{eqnarray}
\label{backg}
F_{\hat{i}\hat{j}}&=&L^{a}_{\hat{i}}E_{ab}L^{b}_{\hat{j}}; \nonumber \\
F_{\hat{i}\beta}&=&L^{a}_{\hat{i}}E_{ab}(E^{0}_{bc})^{-1}F^{R}_{c\beta};
\nonumber \\
F_{\alpha\hat{j}}&=&F^{L}_{\alpha a}(E^{0}_{ab})^{-1}E_{bc}L^{c}_{\hat{j}};
\nonumber \\
F_{\alpha\beta}&=&\hat{F}_{\alpha\beta}+
F^{L}_{\alpha a}((E^{0})^{-1}E(E^{0})^{-1}-(E^{0})^{-1})^{ab}F^{R}_{b\beta}.
\end{eqnarray}
The dual theory is found by decomposing $L$ as $V\tilde{U}$. When we 
integrate out $V$ again by changing variables we find the background
\begin{eqnarray}
\label{dualbackg}
\tilde{F}^{\hat{i}\hat{j}}&=&\tilde{L}^{\hat{i}}_{a}\tilde{E}^{ab}
\tilde{L}^{\hat{j}}_{b}; \nonumber \\
\tilde{F}^{\hat{i}}_{~\beta}&=&\tilde{L}^{\hat{i}}_{a}\tilde{E}^{ab}
F^{R}_{b\beta};
\nonumber \\
\tilde{F}_{\alpha}^{~\hat{j}}&=&-F^{L}_{\alpha a}\tilde{E}^{ab}
\tilde{L}^{\hat{j}}_{b};
\nonumber \\
\tilde{F}_{\alpha\beta}&=&\hat{F}_{\alpha\beta}-
F^{L}_{\alpha a}\tilde{E}^{ab}F^{R}_{b\beta}.
\end{eqnarray}
The generalized Buscher transformation are
\begin{eqnarray}
\label{genBT}
\tilde{E}^{-1}-\beta\alpha=(E^{-1}+\mu\nu)^{-1}&=&E^{0}(x^{\alpha}); 
\nonumber \\
\tilde{E}^{-1}\tilde{\cal F}^{R}=E^{0}E^{-1}{\cal F}^{R}&=&F^{R}; \nonumber \\
-\tilde{{\cal F}}^{L}\tilde{E}^{-1}={\cal F}^{L}E^{-1}E^{0}&=&F^{L};
\nonumber \\
\tilde{F}-\tilde{\cal F}^{L}\tilde{E}^{-1}\tilde{\cal F}^{R}=
F+{\cal F}^{L}(E^{-1}E^{0}E^{-1}-E^{-1}){\cal F}^{R}&=&\hat{F},
\end{eqnarray}
where ${\cal F}_{\alpha b}^{L}\equiv F_{\alpha\hat{j}}L^{\hat{j}}_{~b}$, 
${\cal F}^{R}_{a\beta}\equiv L_{a}^{~\hat{i}}F_{\hat{i}\beta}$, 
$\tilde{{\cal F}}_{\alpha}^{Lb}\equiv 
\tilde{F}_{\alpha}^{~\hat{j}}\tilde{L}_{\hat{j}}^{~b}$ and 
$\tilde{{\cal F}}^{Ra}_{~~\beta}\equiv 
\tilde{L}^{a}_{~\hat{i}}\tilde{F}^{\hat{i}}_{~\beta}$. 
The bosonic version of these generalized Buscher transformation was given in 
this form in \cite{kn:TvU} and, earlier in a different form in \cite{kn:KS}. 
In the abelian case where $\beta=\mu=0$ a combination of 
eq.(\ref{backg}) and (\ref{dualbackg}) gives eq.(\ref{abelianBT}).

We end this section by showing how the ${\cal N}=1$ dilaton superfield $\Phi$ 
transforms under PL duality. Since the transformation of $\Phi$
is analogous to the transformation of the bosonic component $\phi$ given in 
eqs.(\ref{dilatontransf1})-(\ref{dilatontransf2}) we just give the 
transformation law for the fermionic component
\begin{eqnarray}
\rho_{\pm}&=&\rho^{0}_{\pm}+M_{U}|\lambda_{\pm}-
(M_{\alpha}|-N_{\alpha}|)\eta^{\alpha}_{\pm}; \nonumber \\
\tilde{\rho}_{\pm}&=&\rho^{0}_{\pm}+\tilde{M}_{\tilde{U}}|\tilde{\lambda}_{\pm}
+\tilde{M}_{\alpha}|\eta^{\alpha}_{\pm}
\end{eqnarray}
and the auxiliary field
\begin{eqnarray}
W&=&W^{0}+iM_{UU}|\lambda_{-}\lambda_{+}+M_{U}|Y
+iM_{\alpha U}|\eta^{\alpha}_{[-}\lambda_{+]}
\nonumber \\
&&+i(M_{\alpha\beta}|-N_{\alpha\beta}|)\eta^{\alpha}_{-}\eta^{\beta}_{+}
-N_{\alpha}T^{\alpha}; \nonumber \\
\tilde{W}&=&W^{0}+
i\tilde{M}_{\tilde{U}\tilde{U}}|\tilde{\lambda}_{-}\tilde{\lambda}_{+}
+i\tilde{M}_{\alpha\tilde{U}}|\eta_{[-}^{\alpha}\tilde{\lambda}_{+]}
+i\tilde{M}_{\alpha\beta}|\eta_{-}^{\alpha}\eta_{+}^{\beta} \nonumber \\
&&+\tilde{M}_{\tilde{U}}|\tilde{Y}+\tilde{M}_{\alpha}|T^{\alpha}.
\end{eqnarray}
Here we have defined $M(U,X^{\alpha})=\ln\det E(U,X^{\alpha})$, 
$N(X^{\alpha})=\ln\det E^{0}(X^{\alpha})$ and $M_U$ means the derivative of
$M$ w.r.t. $U$ and $N_{\alpha}$ means the derivative of $N$ w.r.t. 
the spectator field $X^{\alpha}$ and $|$ means ``the $\theta$-independent 
component of''. 
The component fields are defined as follows:
$\lambda_{\pm}=D_{\pm}U|$, $Y=D^{2}U|$, $\rho_{\pm}=D_{\pm}\Phi|$, 
$W=D^{2}\Phi|$, $\eta^{\alpha}_{\pm}=D_{\pm}X^{\alpha}|$ and 
$T^{\alpha}=D^{2}X^{\alpha}|$. The dual components are defined similarly.

\section{${\cal N}=1$ non-abelian duality}
\label{sectSNAD}

In this section we give an explicit representation of the generators of the
Drinfel'd double relevant for non-abelian duality. We also discuss
non-abelian duality for ${\cal N}=1$ supersymmetric models in the
PL path-integral setting, extending the bosonic analysis 
\cite{kn:TvU}.

In the PL setting non-abelian duality corresponds to the dual group 
$\tilde{G}$ being abelian \cite{kn:KS}. Correspondingly we choose a Drinfel'd 
double $G\otimes U(1)^{dimG}$, where $\{T_{a}\}$ span ${\cal G}$ and 
$\{T^{a}\}$ span $\tilde{\cal G}=\sum_{1}^{dimG}\oplus u(1)$; 
$(a=1,\ldots,\mbox{dimG})$ and the
two sets of generators satisfy the algebra (ref. eq.\ref{algebra})
\begin{equation}
[T_{a},T_{b}]=f_{ab}^{c}T_{c};\ \ \
[T^{a},T^{b}]=0;\ \ \
[T_{a},T^{b}]=-f_{ac}^{b}T^{c}.
\end{equation}
An explicit representation of the set $\{T_{a},T^{a}\}$ is
\begin{equation}
T_{a}=\left(\begin{array}{cc}
                t_{a} & 0 \\[1mm]
                0 & t_{a}  \\
                \end{array}
\right),~~~T^{a}=\left(\begin{array}{cc}
                0  & \frac{1}{\lambda}t_{a} \\[1mm]
                0 & 0 \\
                \end{array}
\right),
\end{equation}
where $t_{a}\in {\cal G}$ satisfy the algebra 
$[t_{a},t_{b}]=f_{ab}^{~~c}t_{c}$ and $Tr(t_{a}t_{b})=\lambda\delta_{ab}$.
Consistency requires that the structure constants are completely
antisymmetric. Thus this representation is possible only when the group $G$ is
compact.

Let $X$ and $Y$ be two $2\times 2$ block matrices where each block is a
$r\times r$ matrix. 
We define the inner product (\ref{innerprod}) as
\begin{equation}
\langle X,Y\rangle=Tr[XY]_{12},
\end{equation}
where $[~~]_{12}$ stands for the one-two block of the product $XY$.
Since $\tilde{f}^{ab}_{~~c}=0$, the bi-Lie algebra consistency condition 
(\ref{consistency}) is trivial fulfilled.

Writing the group element of $G$ as $\exp(X^{a}T_{a})$ (and a similar 
parametrization of the dual group elements), we find
\begin{equation}
U=\left(\begin{array}{cc}
                u & 0 \\[1mm]
                0 & u  \\
                \end{array}
\right),~~~\tilde{U}=\left(\begin{array}{cc}
                \bf{1} & \frac{1}{\lambda}\tilde{X}_{a}t_{a} \\[1mm]
                0 & \bf{1} \\
                \end{array}
\right),
\end{equation}
where $\bf{1}$ is the $r\times r$ identity matrix and 
$u=\exp(X^{a}t_{a})\in G$.
Moreover we find
\begin{eqnarray}
\label{relations}
\mu^{ab}(U)&=&\mu^{ab}(U^{-1})=0;\ \ \ \ \
\nu^{a}_{~b}(U)=\nu^{b}_{~a}(U^{-1})=
\frac{1}{\lambda}Tr(ut_{a}u^{-1}t_{b}); \nonumber \\
\alpha_{a}^{~b}(\tilde{U})&=&\alpha_{a}^{~b}(\tilde{U}^{-1})
=\delta_{ab};\ \
\beta_{ab}(\tilde{U})=-\beta_{ab}(\tilde{U}^{-1})=\tilde{X}_{c}f^{b}_{ca}
=f_{ab}^{~~c}\tilde{X}_{c}
\end{eqnarray}
From these relations it is easy to find the backgrounds
\begin{eqnarray}
E_{ab}(u)=E^{0}_{ab};\ \ \ \ \
\tilde{E}^{ab}(\tilde{u})=[(E^{0}+f^{c}\tilde{X}_{c})^{-1}]^{ab},
\end{eqnarray}
where $f^{c}\tilde{X}_{c}\equiv f^{~~c}_{ab}\tilde{X}_{c}$. The ${\cal N}=1$
non-abelian Buscher transformation can be found from eq.(\ref{genBT}) by
setting $E=E^{0}$, $\mu=0$ and $\alpha={\bf 1}$.

We now turn to a comparison to the usual non-abelian duality formulation.
A ${\cal N}=1$ supersymmetric generalizing of 
(\ref{I+Lagrmult}), (for simplicity we will set the spectators to zero in this 
section) is
\begin{equation}
\label{conAsusy}
I[L,C_{-}]=I[L]+i\int d^{2}\xi d^{2}\theta
\left[\langle L^{-1}D_{+}L,T^{a}\rangle E^{0}_{ab}
-\langle L^{-1}D_{+}L,T_{b}\rangle\right]C_{-}^{b},
\end{equation}
where $C_{-}^{b}$ is a Lagrange multiplier. From (\ref{conAsusy}) we may 
recover the ${\cal N}=1$ formulation of the 
traditional non-abelian duality in the form it has after gauge-fixing 
the coordinates on $G$ to zero. Decomposing $L=V\tilde{U}\in D$, $V\in G$, 
$\tilde{U}\in \tilde{G}$, the action becomes
\begin{equation}
\label{gfixedaction}
I[A_{+},\tilde{J}_{\pm},\tilde{X},C_{-}]=i\int d^{2}\xi d^{2}\theta
[A^{a}_{+}(\alpha^{-1})_{a}^{~b}\tilde{J}_{-b}+
(A^{a}_{+}(\alpha^{-1})_{a}^{~c}E_{cb}^{0}
-A^{a}_{+}\beta_{ab}^{t}-\tilde{J}_{+b})C_{-}^{b}],
\end{equation}
where $A_{+}\equiv V^{-1}D_{+}V$ and 
$\tilde{J}_{\pm}\equiv\tilde{U}^{-1}D_{\pm}\tilde{U}$.
After using eq.(\ref{relations}), and using the abelian current 
$\tilde{J}_{\pm a}=D_{\pm}\tilde{X}_{a}$, the action 
(\ref{gfixedaction}) may be written
\begin{eqnarray}
\label{susynaaction}
I[A_{+},C_{-},\tilde{X}_{a}]=i\int d^{2}\xi d^{2}\theta [
(A^{a}_{+})D_{-}\tilde{X}_{a}-(D_{+}\tilde{X}_{a})C_{-}^{a}+
A_{+}^{a}(E_{ab}^{0}+f^{c}_{ab}\tilde{X}_{c})C_{-}^{b}]. 
\end{eqnarray}
This is the first order action (up to a total derivative) one usually starts 
with to obtain the dual model in the ${\cal N}=1$ non-abelian duality.
The action (\ref{susynaaction}) needs some comments. First of all the 
components of the gauge field are $A^{a}_{+}$ and $C^{a}_{-}$, where 
$C^{a}_{-}$ as we remember was from the beginning a Lagrange multiplier. The 
``new'' Lagrange multiplier is $\tilde{X}_{a}$ \cite{kn:TvU}. Note that in 
the abelian limit the terms involving the Lagrange multiplier
\begin{equation}
i\int d^{2}\xi d^{2}\theta [
(A^{a}_{+})D_{-}\tilde{X}_{a}-(D_{+}\tilde{X}_{a})C_{-}^{a}]=
i\int d^{2}\xi d^{2}\theta\tilde{X}_{a}F^{a} + \mbox{surface term}, 
\end{equation}
where $F^{a}=D_{-}A^{a}_{+}+D_{+}C_{-}^{a}$.
This is exactly the correct term one needs to be sure to get back the original
action when the Lagrange multiplier is integrated out \cite{kn:GR1,kn:GR2}. 

After a partial integration and variation of $I$ w.r.t. $\tilde{X}$, 
we get the zero ``field strength'' condition
\begin{equation}
\label{zeroF}
D_{-}A^{a}_{+}+D_{+}C_{-}^{a}+A_{+}^{b}f^{a}_{bc}C_{-}^{c}=0.
\end{equation}

This is the condition that may be solved to give back the original model.

\section{Discussion}

In this paper we have presented a path integral formulation of ${\cal N}=1$ 
supersymmetric PL duality. We have shown that it arises as a straightforward 
generalization of the treatment of the bosonic case. In this context we have 
also given an explicit representation of the generators of the Drinfel'd 
double bi-algebra corresponding to the group $G\times U(1)^{dimG}$ relevant 
for non-abelian dualization, and used it to elucidate the relation between the
actions describing PL duality and the parent action in non-abelian 
${\cal N}=1$ supersymmetric duality. 

With applications to the ${\cal N}=2$ supersymmetric case in mind, we have 
also touched upon WZW models. In particular, in Appendix \ref{appB} we show 
how the treatment of the WZW model in \cite{kn:GR2} yields a 
formulation open also to PL duality. 

The ${\cal N}=2$ supersymmetric case presents considerable difficulty, 
however. This is due to the severe constraints on the target manifold that 
such $\sigma$-models imply \cite{kn:GHR}. Our best approach so far to 
this problem is via an ${\cal N}=1$ superspace formulation where we may make 
use of the results in the present paper. Several questions such as the 
transformations of the complex and direct product structures under PL duality 
remain open however, and we have not found a complete characterization of the 
duality as yet. We hope to return to the question in a later publication.

The question of whether a PL type duality can be based on the extended 
currents that result when the Lie derivative of the background antisymmetric 
tensor field $b$ satisfies ${\cal L}b=d\omega$ may seem natural at a first 
encounter with PL duality. In Appendix \ref{appA} we discussed this question a
little and found that under certain very special circumstances this can indeed
be done. The answer is still essentially negative, though, high-lighting the 
special status of PL duality.

\begin{flushleft} {\bf Acknowledgments}
\end{flushleft}

\noindent We thank Rikard von Unge for discussions.
UL gratefully acknowledges the hospitality of Oslo University, support
by NFR/grant 650-1998368 and EU contract HPRN-CT-2000-0122.

%\newpage

\appendix

\section{A Modified Current and Poisson-Lie\\ Duality}
\label{appA}

In this appendix we discuss some aspects of PL duality with modified 
currents.

Instead of isometries $k_a$ that also preserve the $b$ field 
${\cal{L}}_{k_a}b_{ij}=0$, we here turn to the case discussed in 
section \ref{sect2}: isometries that satisfy 
${\cal{L}}_{k_a}b_{ij}=\partial_{[i}\omega_{j]a}$, 
(\ref{iso-cond2}). The question we address is whether it is 
possible to find a Maurer-Cartan equation as in (\ref{Maurer1}), but for 
currents corresponding to (\ref{mod-curr}) rather than (\ref{curr}).

We first note the following way of writing the Lie derivative of the $b$ 
field:
\begin{equation}
\label{lder} 
{\cal{L}}_{k_a}b_{ij}=k_a^{k}H_{ijk}-\partial_{[i}(k_a^kb_{j]k}).
\end{equation}
This shows that when $k_a$ represent an invariance of the action and thus
$k_a^kH_{ijk}=\partial_{[i}v_{j]a}$ (see below (\ref{varyWZW})), we may set
\begin{equation}
\label{omdef}
\omega_{ia}=v_{ia}-k_a^kb_{ik},
\end{equation}
This ensures integrability, i.e., 
$[{\cal{L}}_{k_{a}},{\cal{L}}_{k_{b}}]b_{ij}=f_{ab}^c{\cal{L}}_{k_{c}}b_{ij}$.

We now turn to the general case and replace $k_a^i\to R_a^i$ as before. With 
this replacement (\ref{lder}) is still valid, but the action is no-longer 
taken to be invariant. We further assume that the $b$-field may be split 
according to
\begin{equation}
b_{ij}=\hat b_{ij}+b^0_{ij},
\end{equation}
where ${\cal{L}}_{R_a}b^0_{ij}=\partial_{[i}\omega^0_{j]a}$,
i.e., where the field strength $H^0_{ijk}$ of $b^0_{ij}$ is preserved 
by $R_a^i$.

The field equations (\ref{condition}) that result from the action 
(\ref{varS}) may then be rewritten as
\begin{equation}
\label{hatfes}
\partial\hat{\bar{J}}_a+\bar{\partial}\hat{J}_a
-{\cal{L}}_{R_a}\hat f_{ij}\partial x^{i}\bar{\partial}x^{j}=0,
\end{equation}
with the currents as in (\ref{mod-curr}),
\begin{equation}
\hat{J}_a=\partial x^i(f_{ij}R_a^j+\omega^0_{ia});\qquad
\hat{\bar{J}}_a=(R_a^if_{ij}-\omega^0_{ja})\bar\partial x^j,
\end{equation}
and 
\begin{equation}
\hat f_{ij}=f_{ij}-b^0_{ij}.
\end{equation}
In the general case we have found no modification of the PL condition 
satisfying the integrability conditions, which can turn the equation 
(\ref{hatfes}) into a Maurer-Cartan equation. Below we remark on a few special
cases.

The first (trivial) case we consider is when $\omega_{ia}^0=0$.  The currents 
are then unmodified $\hat J_a=J_a$ and the PL condition is also 
unmodified
\begin{equation}
{\cal{L}}_{R_a}\hat f_{ij}={\cal{L}}_{R_a} f_{ij}=
- f_{ik}R^k_b\tilde f^{bc}_aR^l_c f_{lj}.
\end{equation}
PL duality thus allows for a part of the $b$-field to be preserved by
the Lie-derivative.

Second, we assume that $v_{ia}^0$ in (\ref{omdef}) may be gauged to zero using
the $b$ gauge transformations (cf comment below (\ref{iso-cond2})). This means 
that $\omega^0_{ia}=R^k_ab^0_{ki}$ and this eliminates all dependence on 
$b^0_{ij}$ in $\hat J$,$\hat{\bar J}$ and in $\hat f$. If we then {\em modify}
the PL condition to read
\begin{equation}
\label{hatpl}
{\cal{L}}_{R_a}\hat f_{ij}=
-\hat f_{ik}R^k_b\tilde  f^{bc}_aR^l_c\hat f_{lj},
\end{equation}
we find the 
Maurer-Cartan equations for the hatted currents:
\begin{equation} 
\partial\hat{\bar{J}}_a+\bar\partial\hat{J}_a
+\hat J_b\tilde f^{bc}_a\hat{\bar J}_c=0.
\end{equation}
(Integrability of (\ref{hatpl}) is ensured as before when $\cal{G}$ and 
$\tilde{\cal{G}}$ form a bi-Lie-algebra as in (\ref{consistency})).
We see that we may indeed use the currents $\hat J$ and $\hat{\bar J}$ as a 
starting point for PL dualization in this case.

Finally we consider the case when $v_{ia}^0=R^k_ag_{ik}$ in (\ref{omdef}). 
When $R_a$ generate isometries and this relation holds for the whole 
$b$-field, the symmetry algebra is an infinite dimensional Kac-Moody algebra 
\cite{kn:HS} and 
the currents reduce to one chiral current \cite{kn:RV}. This latter fact 
results even if we do not assume isometries, since 
$\omega^0_{ia}=R^k_af^0_{ki}$ which gives 
\begin{equation}
\hat J_a=\partial x^i\left(2g_{ij}+(b-b^0)_{ij}\right)R^j_a,\quad
\hat{\bar J}_a=-(b-b^0)_{ij}R^j_a\bar{\partial }x^i.
\end{equation}
Clearly $J_a = 2g_{ij}\partial x^iR^j_a$ and  $\bar J_a$ 
vanishes when $b_{ij}=b_{ij}^0$. In this case the PL condition 
applied to $f_{ij}$ will lead to a PL dual model (expressed in 
$J,\bar{J}$, whereas 
$\hat f_{ij}=g_{ij}$ implies that the modified condition (\ref{hatpl}) 
vanishes (thus it only holds if we do have an isometry).

\section{A comment on WZW-models and Poisson-Lie Duality}
\label{appB}

The WZW model on $G$ given in eq.(\ref{wzw}) is not on the form 
used as starting point in the PL dualization scheme, because it 
is not on the form (\ref{boseaction2}). It follows from (\ref{backgr}) that
\begin{equation}
b_{ij}=\frac{1}{2}L^{a}_{[i}E_{ab}L^{b}_{j]}.
\end{equation}
Hence writing (\ref{wzw}) in the form (\ref{boseaction2}), given 
(\ref{wzw-term}), requires the relation 
\begin{equation}
\partial_{[i}(L_{j}^{a}E_{ab}L_{k]}^{b})=
\frac{1}{2}f_{abc}L_{i}^{a}L_{j}^{b}L_{k}^{c},
\end{equation}
to be satisfied. This leads to difficulties, at least for PL formulation of
non-abelian duality (without spectators) where the $E_{ab}$'s must be 
constants.

In \cite{kn:GR2} when discussing non-abelian duality the
following treatment of the WZW-model was given. Here we show how to adopt it 
for PL duality. Consider the $WZW$-model on $G\times G$. 
The action is written \cite{kn:GR2}
\begin{equation}
S_{G_{k}\times G_{k}}[g_1,g_2]=S_{G_{k}}[g_1]+S_{G_{k}}[g_2],
\end{equation}
where $S_{G_{k}}[g]=(k/2\pi)S[g]$; here $S[g]$ is defined in eq.(\ref{wzw}). 
Next we define two new variables
\begin{equation}
g\equiv g_{1};\ \ \ \ \ x\equiv g_{2}g_{1}.
\end{equation}
The element $x$ is inert under the (gauge) transformation 
$g_{1}\rightarrow u^{-1}g_{1}$, 
$g_{2}\rightarrow g_{2}u$. 
Using the Polyakov-Wiegmann formula 
\begin{equation}
S_{G_{k}}[g_{1}g_{2}]=S_{G_{k}}[g_{1}]+S_{G_{k}}[g_{2}]+
\frac{k}{2\pi}\int d^2\xi Tr(g_{1}^{-1}\partial g_{1}
\bar{\partial}g_{2}g_{2}^{-1})
\end{equation}
and that the $WZW$ term satisfies $\Gamma[g^{-1}]=-\Gamma[g]$ we rewrite
the action $S_{G_{k}\times G_{k}}$ as
\begin{equation}
\label{wzwk}
S_{G_{k}\times G_{k}}[g,x]
=\frac{k}{2\pi}\int d^{2}\xi Tr(g^{-1}\partial g g^{-1}\bar{\partial}g
-x^{-1}\partial x g^{-1}\bar{\partial} g)+S_{G_{k}}[x],
\end{equation}
where $S_{G_{k}}[x]$ is an ordinary $\sigma$-model on a manifold
with background $f_{\alpha\beta}(x)$.
The action (\ref{wzwk}) can be compared with the original action found by 
integrating out the Lagrange multipliers $\lambda_{a}$ of the action 
(\ref{firstorder}). 
Remembering the correspondence 
$x^{-1}\partial x=t_{a}L^{a}_{\alpha}\partial x^{\alpha}$
we see that 
$E_{ab}\equiv L^{\hat{i}}_{a}f_{\hat{i}\hat{j}}L^{\hat{j}}_{b}=
\frac{k}{2\pi}Tr(t_{a}t_{b})$, $F^{L}_{\alpha b}=
-\frac{k}{2\pi}L_{\alpha}^{~a}Tr(t_{a}t_{b})$,
$F_{a\beta}^{R}=0$.

\newpage
%\vskip 0.5in


\baselineskip=1.6pt
\begin{thebibliography}{99}
\bibitem{kn:KS} 
C. Klim\v{c}ik and P. \v{S}evera, 
{\em Dual Non-Abelian Duality and the Drinfeld Double}, 
Phys. Lett. {\bf B 351} (1995) 455 [hep-th/9502122]; 
C. Klim\v{c}ik, 
{\em Poisson-Lie T-Duality}, 
Nucl. Phys. {\bf 46} (Proc. Suppl.) (1996) 116 [hep-th/9509095].
\bibitem{kn:AKT}
A. Y. Alekseev, C. Klim\v{c}ik and A. A. Tseytlin,
{\em Quantum Poisson-Lie $T$-duality and WZNW model},
Nucl. Phys. {\bf B 458} (1996) 430 [hep-th/9509123].
\bibitem{kn:TvU} 
E. Tyurin and R. von Unge, 
{\em Poisson-Lie T-Duality: the Path-Integral Derivation}, 
Phys. Lett. {\bf B 382} (1996) 233 [hep-th/9512025].
\bibitem{kn:S1}
K. Sfetsos,
{\em Poisson-Lie $T$-duality beyond the classical level and the 
renormalization group},
Phys. Lett. {\bf B 432} (1998) 365 [hep-th/9803019].
\bibitem{kn:P} 
S. E. Parkhomenko,
{\em Poisson-Lie $T$-duality in $N$=2 superconformal field theories},
Sov. Phys. JETP Lett. {\bf 64} (1996) 877 [hep-th/9612034]; 
{\em Poisson-Lie $T$-duality and $N$=2 superconformal $WZNW$
models on compact groups}, Mod. Phys. Lett. {\bf A 12} (1997) 3091 
[hep-th/9705233]; 
{\em Poisson-Lie $T$-duality and complex geometry in $N$=2 superconformal 
$WZNW$ models}, Nucl. Phys. {\bf B 510} (1998) 623 [hep-th/9706199];
{\em Mirror symmetry as a Poisson-Lie $T$-duality}, 
Mod. Phys. Lett. {\bf A 13} (1998) 1041 [hep-th/9710037];
{\em On the quantum Poisson-Lie $T$-duality and mirror symmetry}, 
J. Exp. Theor. Phys. {\bf 89} (1999) 5 [hep-th/9812048].
\bibitem{kn:S2}
K. Sfetsos,
{\em Poisson-Lie $T$-duality and supersymmetry},
Nucl. Phys. {\bf 56B} (Proc. Suppl.) (1997) 302 [hep-th/9611199].
\bibitem{kn:K}
C. Klim\v{c}ik,
{\em Poisson-Lie $T$-duality and $(1,1)$ supersymmetry},
Phys. Lett. {\bf B 414} (1997) 85 [hep-th/9707194].
\bibitem{kn:KP}
C. Klim\v{c}ik and S. Parkhomenko,
{\em Supersymmetric gauged $WZNW$ models as dressing cosets},
Phys. Lett. {\bf B 463} (1999) 195 [hep-th/9906163].
\bibitem{kn:AL}
F. Assaoui and T. Lhallabi,
{\em Supersymmetric quantum corrections and Poisson-Lie $T$-duality},
Class. Quant. Grav. {\bf 18} (2001) 277 [hep-ph/0007337].
\bibitem{kn:RV}
M. Ro\v{c}ek and E. Verlinde, {\em Duality, quotients, and currents}, 
Nucl. Phys. {\bf B 373} (1992) 630 [hep-th/9110053].
\bibitem{kn:GR1}
A. Giveon and M. Ro\v{c}ek, 
{\em Introduction to Duality}, hep-th/9406178.
\bibitem{kn:GR2}
A. Giveon and M. Ro\v{c}ek, {\em On Nonabelian Duality}, 
Nucl. Phys. {\bf B 421} (1994) 173 [hep-th/9308154].
\bibitem{kn:HS}
C. M. Hull and B. Spence, 
{\em The Geometry of the Gauged Sigma-Model with Wess-Zumino Term},
Nucl. Phys. {\bf B 353} (1991) 379;
{\em The Gauged Nonlinear Sigma Model with 
Wess-Zumino Term},
Phys. Lett. {\bf B 232} (1989) 204.
\bibitem{kn:FQ}
X. C. de la Ossa and F. Quevedo,
{\em Duality symmetries from non-abelian isometries in string theory},
Nucl. Phys. {\bf B 403} (1993) 377 [hep-th/9210021];
F. Quevedo,
{\em Abelian and non-abelian dualities in string backgrounds},
hep-th/9305055.
\bibitem{kn:BT} 
T. H. Buscher, 
{\em Path-Integral Derivation of Quantum Duality in Nonlinear Sigma-Models}, 
Phys. Lett. {\bf B 201} (1988) 466.
\bibitem{kn:D}
V.G. Drinfel'd,
{\em Quantum Groups},
in Proc. ICM, MSRI, Berkeley, 1986.
\bibitem{kn:FG}
F. Falceto and K. Gawedzki,
{\em Lattice Wess-Zumino-Witten Model and Quantum Groups},
J. Geom. Phys. {\bf 11} (1993) 251 [hep-th/9209076].
\bibitem{kn:AM}
A. Y. Alekseev and A. Z. Malkin,
{\em Symplectic structures associated to Lie-Poisson groups},
Commun. Math. Phys. {\bf 162} (1994) 147 [hep-th/9303038].
\bibitem{kn:PW}
A. M. Polyakov and P. B. Wiegmann,
{\em Theory of nonabelian Goldstone bosons in two dimensions},
Phys. Lett {\bf B 131} (1983) 121.
\bibitem{kn:HKLR}
C. M. Hull, A. Karlhede, U. Lindstr\"{o}m and M. Ro\v{c}ek,
{\em Nonlinear $\sigma$-models and their gauging in and out of superspace},
Nucl. Phys. {\bf B 266} (1986) 1.
\bibitem{kn:GHR}
S. J. Gates Jr., C. M. Hull and M. Ro\v{c}ek,
{\em Twisted Multiplets and New Supersymmetric Non-Linear $\sigma$-Models},
Nucl. Phys. {\bf B 248} (1984) 157;
P. S. Howe and G. Sierra,
{\em Two-Dimensional Supersymmetric Nonlinear $\sigma$-Models with Torsion},
Phys. Lett. {\bf B 148} (1984) 451;
M. Ro\v{c}ek, 
{\em Modified Calabi-Yau Manifolds with Torsion},
in {\em Mirror symmetry I}, Yau, S. T. ed., AMS International Press, 
Cambridge 1998.
\end{thebibliography}
\end{document}